\documentclass[conference]{IEEEtran}

\usepackage[normalem]{ulem} % for \sout
\usepackage{xcolor,xspace}

\usepackage{ifthen}
\usepackage{amssymb}
\newboolean{showcomments}
\setboolean{showcomments}{true} % toggle to show or hide comments
\ifthenelse{\boolean{showcomments}}
{\newcommand{\nb}[2]{
		\fcolorbox{gray}{yellow}{\bfseries\sffamily\scriptsize#1}
		{$\blacktriangleright$#2$\blacktriangleleft$}
	}
	
}
{\newcommand{\nb}[2]{}
	
}

\usepackage{url}
\usepackage{hyperref}
\usepackage{epsfig}
\newcommand{\eg}{e.g.,~}										% exempli gratia (for the sake of example)
\newcommand{\ie}{i.e.,~}										% id est (that is)
\newcommand{\Fig}[1]{Fig.~\ref{#1}}  			% choose Fig. or Figure, depending on the style
\newcommand{\Sect}[1]{Section~\ref{#1}}	  	% section name always with a capital S
\begin{document}

\title{Automatically Installing and Deploying Tools for Conducting Systematic Reviews in ReLiS}

\author{
	\IEEEauthorblockN{Eugene Syriani and Brice M. Bigendako}
	\IEEEauthorblockA{University of Montreal\\
		syriani@iro.umontreal.ca, brice-michel.bigendako@umontreal.ca}
}

\maketitle

\begin{abstract}
Conducting systematic reviews (SR) is a time consuming endeavor that requires several iterations to setup right.
We present ReLiS, a framework to configure and deploy projects while conducting a SR.
It features a domain-specific modeling editor tailored for researchers who perform SRs and an architecture that enables live installation and deployment  of multiple concurrently running projects. See the accompanying video at \url{http://youtu.be/U5zOmk2vWy8}
\end{abstract}

\section{Introduction}

One of the key benefits of model-driven development (MDD) is the automatic synthesis of code from domain-specific models~\cite{Mohagheghi2008a}.
In most reported applications, the generated code is typically part of the core functionality of the software product.
However, the generation of installers and automatic deployment has not yet been addressed in the MDD community.
This is especially useful in the domain of web applications.
Base platforms to build web applications (\eg Joomla, WordPress) support easy installation of custom extensions.
However, their development requires the user to install the appropriate tooling and programming environment.
Also, they are very generic and provide too many options that are not used in specific business application domains.

In this paper, we concentrate on the area of evidence-based software engineering, in particular on secondary studies such as systematic literature reviews (SLR) and mapping studies (SMS)~\cite{Kitchenham2004}.
When a researcher desires to address a specific research problem, he starts by looking at what already exists in the literature on the topic.
Therefore, these kinds of reviews are used by researchers in many areas of science and engineering~\cite{Thomas2010}.
Conducting large SRs requires considerable effort~\cite{Petersen2008}.
Therefore, tools~\cite{Marshall2013} to help manage references, perform screening, collect meta-data, and report results of the study are of tremendous value for researchers.

The research challenge this paper addresses is the ability to automatically install and deploy complex web-based applications seamlessly.
We employ a model-driven approach which does not impose any prior programming skill from the user.
To demonstrate the usefulness of our approach in the domain of SR, we have developed the framework ReLiS\footnote{
	ReLiS stands for ``Revue Litt{\'e}raire Syst{\'e}matique'' which is French for SLR. ``Relis'' literally translates to ``reread''.},
which offers an integrated environment for installing and configuring SR projects so that a group of researchers can conduct a SR collaboratively using an application customized to their desired research topic.
The contribution of ReLiS is the holistic solution to not only manage the SR process by tracking progress and storing related data, but also in the live installation and configuration of SR projects.

In \Sect{sec:slr}, we briefly outline how SRs are conducted in ReLiS.
In \Sect{sec:mde}, we describe the architecture that enables live installation of online SR projects in ReLiS.
Then in \Sect{sec:dsl}, we explain how we use domain-specific modeling and MDD to define configurations for SR projects, making their installation seamless to the user.
In \Sect{sec:eval}, we present a preliminary evaluation of ReLiS.
Finally, in \Sect{sec:rw} we outline some related work and conclude in \Sect{sec:conclusion}.

\section{A Tool for Conducting Systematic Reviews}\label{sec:slr}

A SLR is a means of evaluating and interpreting all available research relevant to a particular research question, topic area, or phenomenon of interest~\cite{Kitchenham2004}.
A SMS provides a wide overview of a research area to establish if research evidence exists on a topic and provide indication of the quantity of the evidence~\cite{Petersen2008}.
Although these two types of secondary studies have different intentions, they share many commonalities in the way they are conducted and underlying tools they rely on.
Without loss of generality, we will consider them both as systematic reviews (SR) and rely on SMS specificities for the purpose of this work.

We briefly outline the typical process of conducting a SR, although many variants exist~\cite{Kitchenham2009}.
First, the researcher identifies a research question or topic and retrieves papers from the scientific literature accordingly.
With exclusion and inclusion criteria he has predefined for his topic, he screens the corpus of papers to only retain the relevant ones.
Then, the researcher extracts the relevant data from each paper according to the categories of a classification scheme he predefined for the study.
Finally, he analyzes the results with different statistical methods and reports them.

Once the initial corpus of papers is available, the researcher can use ReLiS to ensure he correctly and systematically follows the review procedure he configures.
To do so, he defines a SR project in ReLiS which gets installed automatically on the web server and database where ReLiS is running---\ie the cloud.
Then, he uploads the meta-information of each paper (\eg title, abstract, venue, author) and a link to the full text in his ReLiS project.
The researcher can start screening the corpus and decide which paper to include and which one to exclude.
There can be more than one screening phase: based on the meta-information, on the abstract, and/or on the full text.
Each paper can be assigned automatically or manually to a number of reviewers.
ReLiS supports different methods to validate the screening, such as randomly selecting a number of screened papers and assigning them to another reviewer.
The user can also to decide on a strategy in case of conflicting decisions between different reviewers of the same paper.
ReLiS automatically generates the data extraction forms that researchers can fill online to classify each paper in different categories.
The tool tracks the progress and reports basic statistics for each phase, rendered as tables and plots that can be exported for further analysis.

\section{Architecture for Live Installation}\label{sec:mde}

\begin{figure}
	\centering
	\includegraphics[width=\linewidth]{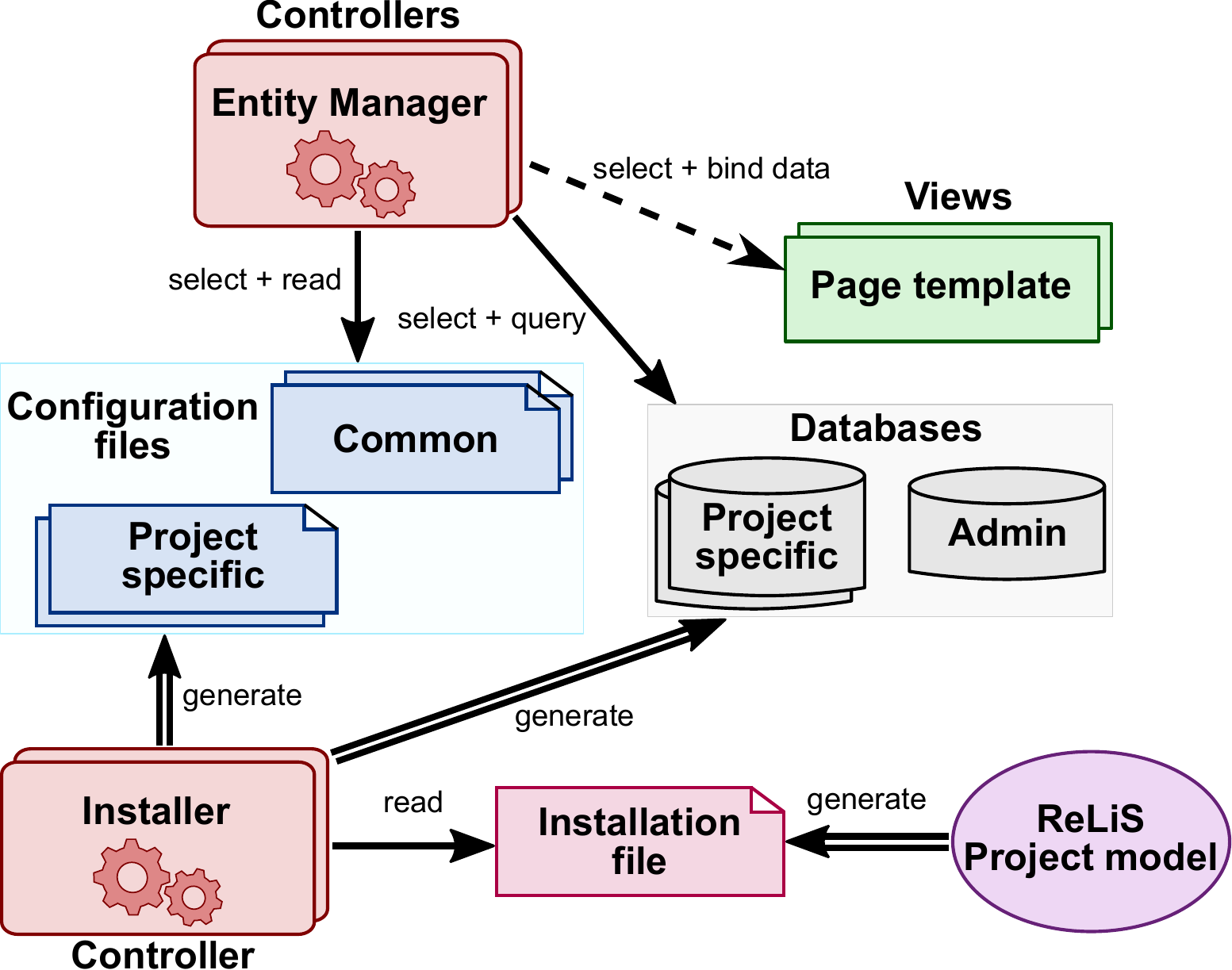}
	\caption{Overview of the architecture to automatically install projects in ReLiS}
	\label{fig:archi}
\end{figure}

Common web content management systems (CMS), \eg Joomla, follow a Model-View-Controller (MVC) architecture.
However, the traditional MVC implemented in CMS has two issues.
As pointed in~\cite{Priefer2016}, this architecture brings a certain overhead and schematically redundant code which requires more effort in manually developing these extensions.
Second, it requires to manually install the extensions and link them to the CMS.
This means that every time one needs to change a functionality that is beyond simple user interface customizations, the user needs to re-install the extension.
To solve these issues, we implemented a dynamic MVC architecture, which is illustrated in the top part of \Fig{fig:archi}.

\subsection{Dynamic MVC architecture}\label{sec:archi}

\begin{figure*}
	\centering
	\includegraphics[trim={0 3.5cm 0 3.6cm},clip,width=\linewidth]{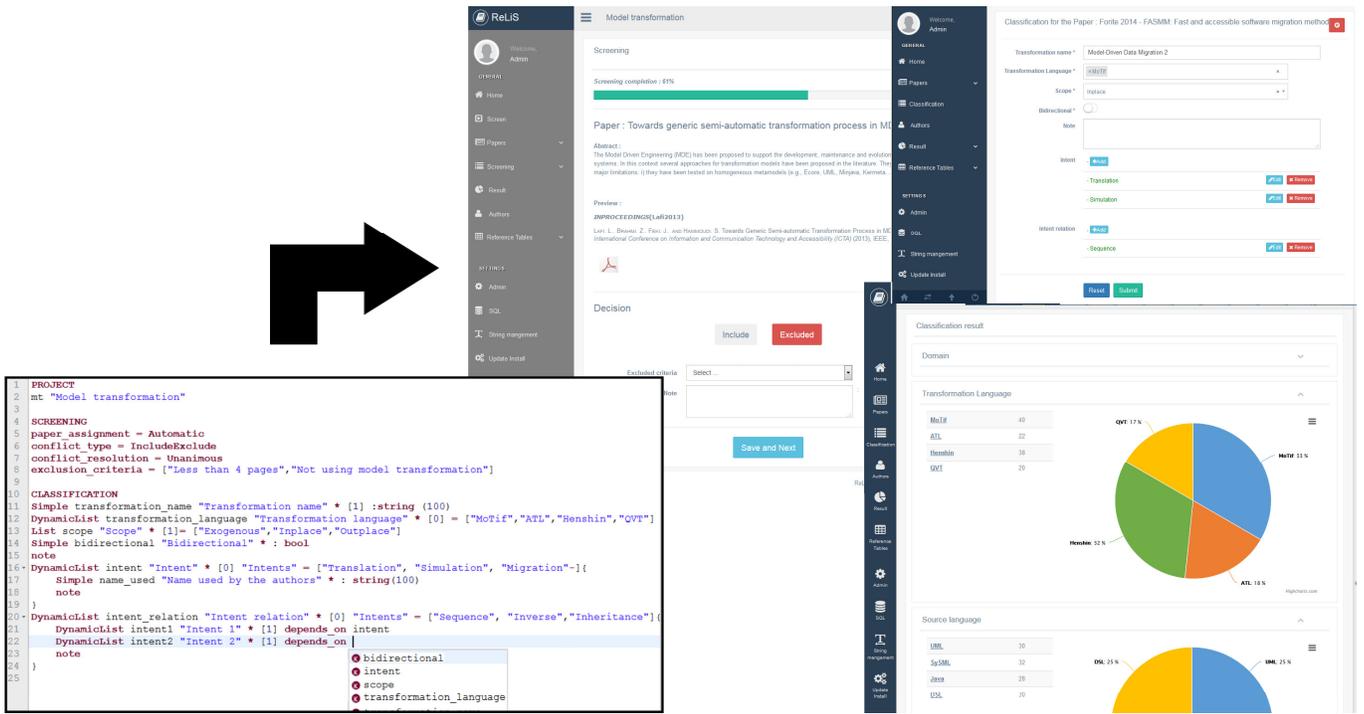}
	\caption{ReLiS model to configure the SR process of a project and how it appears after installing it}
	\label{fig:config}
\end{figure*}

Instead of a model, a view and a controller, ReLiS relies on a collection of each, encapsulated in factories.
There is one \textit{controller} per type of operation.
A particular type of controller is dedicate to managing entities: adding, modifying, removing, listing, or viewing the details of an entity.
For example, as illustrated on the right side of \Fig{fig:config}, a controller is responsible for viewing the details of the meta-data of a paper, another one for modifying the classification of a paper, and a last one for listing the results of the classification.
There are other controllers dedicated to specific operations, such as the installation of projects and user authentication.
Controllers are implemented as generic PHP classes, making them independent from the entities they operate on: the same controller class can list all papers and all authors.

\textit{Entities} (\ie the model component of MVC) are stored in databases which controllers query.
There is one built-in administration database that stores users, projects and the roles they play in different projects.
All other databases are each specific to individual SR projects and contain entities such as papers, authors, and classification categories.

\textit{Views} represent web page templates which controllers bind data to.
As for controllers, views are also independent from the entity to render.
For example, the same web page template is used for adding a user and adding a paper.

What makes this MVC architecture dynamic is that the semantics of controllers is defined in PHP configuration files.
A \textit{configuration} specifies the type of entity to treat with its attributes, the stored procedures to invoke, and the web pages to render with the navigation logic.
The specific controller and configuration are selected from the context of the HTTP request received.

To maximize reuse and modularity, the architecture of ReLiS contains components common to all projects and components specific to individual projects.
Common configurations, which are built-in ReLiS, include for example how papers and users are configured.
Project specific configurations include how the classification is configured.
Although this distinction is useful conceptually, the architecture does not distinguish them and treats them alike.
This facilitates the integration of new configurations and databases without having to modify any of the existing code.
Furthermore, the advantage of using an interpreted language like PHP prevents the need to re-compile every time a new component is added.

\subsection{Live installation of ReLiS projects}\label{sec:deployment}

ReLiS is able to host multiple SR projects operated concurrently.
One particular operation is the installation of projects.
Internally, a project is defined in a PHP file, called an \textit{installation}, that defines entity types and their attributes.
The installation characterizes what information will be rendered on the screen, constrains the domain of each attributes, and defines information relevant to the database.

The \textit{installer} is a controller that generates the configuration file corresponding to the installation it reads in.
It also creates a new database for the project, with tables and columns corresponding to entities, and the necessary stored procedures.

ReLiS allows a user to make modifications to a project while he (or other users) is operating it.
The presented architecture makes it relatively easy to add new entities or attributes.
However, deletion may corrupt the data already present in the application.
The policy we adopted is to only delete elements from a project if they do not contain any data yet.
Otherwise, the element is deactivated but the data it contains is still retained in the database.
This also allows to revert certain actions in case erroneous manipulation.
Edition of elements, such as renaming, are considered as deletions and additions.
Nevertheless, we could detect changes that do not corrupt the data, such as augmenting the size of a string attribute.
However, performing such detections strongly couples ReLiS to the underlying database used which hampers its portability.

\section{Model-driven configuration of Systematic Review Procedures}\label{sec:dsl}

Our goal is to give the flexibility to users to install projects by themselves without prior system administration skills.
This is particularly useful for SR tools, since the parts of the procedure to follow when conducting an SR (\eg exclusion criteria, classification) is often revised during the process.
This is why we opted for a model-driven approach by defining a domain-specific modeling language (DSL), which allows the user to manipulate models at the level of abstraction of the application domain: \ie SRs.
To support this (possibly iterative) process, users can customize the SR procedure in a controlled way.
The steps common to any SR procedure is built-in in ReLiS, with some level of customization, such as the sequence of screening phases.
Furthermore, some steps are specific to the particular SR project and to the research question or area, \eg the classification scheme.
The DSL is geared specifically to state how the components specific to SR projects shall be configured.

A project configuration model in ReLiS has three parts, as illustrated on the left side of \Fig{fig:config}.
The first part identifies the \textit{project} name and user roles required, such as reviewer, senior reviewer or project administrator.
The second part is specific to the \textit{screening} phase.
The user decides how papers are assigned to reviewers (\eg automatically, minimum number of reviewers per paper).
He states what happens in case of conflicting decisions for a paper's inclusion in the study and lists the different reasons to choose from when a paper is to be excluded.
He can also specify the method for validating the screening process, \eg by forcing a senior review to verify 20\% of the papers excluded.

The last part of the model configures the \textit{data extraction form}.
In the model of \Fig{fig:config}, this corresponds to classification scheme of a SMS.
There are three main types of data categories to collect.
Simple categories allow users to freely enter a value, subject to a specific type (text, boolean choice, numerals, etc.) or to some additional constraints (size, regular expression, etc.).
The first category in the example collects the name of a model transformation as a string of at most 100 characters.
List categories offer to choose a value from a predefined list.
The third category collects the scope of the transformation which can be one of three choices.
Finally, the dynamic list category is similar to the list category, but the user can modify the predefined values.
More category features are also supported, such as: sub-categories within others (between $\{ \ldots \}$), mandatory categories (with an~$^*$), multi-valued categories ($\left[ 0 \right]$ for unlimited amount), and dependency constraints across categories to drill down possible values.

We implemented the DSL with a textual syntax using Xtext~\cite{Bettini2013}.
The model editor is completely integrated in ReLiS thanks to DSLFORGE~\cite{Lajmi2014} which generates online lightweight text editors from an Ecore metamodel and an Xtext grammar.
To install projects in ReLiS, we developed a template-based code generator in Xtend~\cite{Bettini2013} to produce the installation file from the model, as illustrated in the bottom part of \Fig{fig:archi}.
This process initiates the installation of the project in ReLiS.
When used by the appropriate controller and configuration file, the pages on the right side of \Fig{fig:config} are rendered corresponding to the specified model.
The presented approach ensures a natural continuum to the user who can operate the tool for its intended purpose (\ie conducting SRs), as well as instantly install new projects or modify and re-install existing ones.

\section{Preliminary Evaluation}\label{sec:eval}

ReLiS is a prototype under development.
Yet, we have performed a preliminary evaluation to assess the usefulness of our approach.
In the context of a graduate course on empirical methods in software engineering, all students were required to conduct a SR on different research topics in software engineering as part of their project.
There were in total 8 groups of 3 to 4 graduate software engineering students who had no previous experience in conducting SRs.

To test ReLiS in the field, the students were asked to use ReLiS for their SR project.
We deployed ReLiS on an Apache Tomcat server\footnote{The prototype used is available at \url{http://relis.iro.umontreal.ca/}.} with a MySQL database.
This allowed us to, first, have test subjects to verify that the tool was functionally correct.
This also gave us the opportunity to assess whether the category types built-in ReLiS are sufficient to cover all 8 SRs.
All groups used every type of category, but none applied advanced constraints on them.

We were also interested in determining how the live integrated installation of projects was put to use.
For that, we extended ReLiS to log every installation of projects while tracking the modifications performed in the project specific model.
We noted that all re-installations were aimed at improving the classification scheme, since the screening procedure should be fixed from the beginning and changing it during the process would hinder the soundness of the SR.
Students found that automatically re-installing their project was very helpful.
This was expected because, during the learning process, they required several iterations before arriving at the desired classification scheme.
The most positive feedback was that they were able to re-install a modified project without loosing the information on the papers they had already classified.
However, we noted a larger amount of deactivated categories than anticipated.
We discovered that it was mainly because of misunderstanding the syntax of the DSL.

%\subsection*{Evaluation plans}

We are planning a field study within a controlled academic environment where reseachers from different disciplines (not only computer science) will conduct SRs using ReLiS.
The study has two purposes.
First, we want to rigorously assess how live installation of projects is used in practice.
Second, we want to determine if the DSL is fit for a broader range of researchers.
We are also planning to quantitatively evaluate the impact on the productivity of researchers when using ReLiS compared to existing SR tools.

\section{Related Work}\label{sec:rw}

The closest work to ours is JooMDD~\cite{Priefer2016}, rapid development environment for extensions of Joomla.
Their also used MDD: they implemented a DSL where models get automatically generated as Joomla extensions, in order to raise the level of abstraction for web extensions development.
However, the specification of JooMDD models is not integrated in Joomla and requires an external IDE to develop it.
%Therefore the installation of extensions and linking them in Joomla needs to be done manually.

%\es{Compare with CMS}
%CMS are very generic frameworks that heavily rely on extensions developed outside the tool's environment.
%ReLiS project models are designed within ReLiS which requires no client-side installation of any tool, as it runs completely in a web browser, and projects are installed automatically.

%JooMDD related work + IFML

There are several tools available to support conducting SRs that were surveyed in~\cite{Marshall2013}.
Some tools~\cite{Thomas2010,Bowes2012} support the whole SR process, including paper retrieval from online databases.
Others focus on text mining techniques to help extract relevant data from papers~\cite{Felizardo2010}.
However, they do not provide the flexibility of modifying the configuration of the SR procedure in a live fashion as ReLiS users can.
%\cite{Hernandes2012}
%\cite{Marshall2013}
%\cite{Marshall2014}
%\cite{Visser2010}
%\cite{Felizardo2010}

\section{Conclusion}\label{sec:conclusion}

In this demo, we present ReLiS, a framework to automatically install SR projects on the cloud.
Its generic and dynamic architecture allows users to install projects during the process of a SR without manual intervention.
A web-based modeling editor specific to SR empowers users to directly configure and install their project by themselves.
Our ultimate goal is to make this tool accessible to non-computer science researchers who desire to conduct a SR.

\bibliographystyle{IEEEtran}
\bibliography{biblio}

\end{document}